\documentstyle[sprocl]{article}

\def\be{\begin{equation}}
\def\ee{\end{equation}}
\def\bea{\begin{eqnarray}}
\def\eea{\end{eqnarray}}
\begin{document}

\title{  THE RELATIVISTIC AND NON RELATIVISTIC QUARK-ANTIQUARK BOUND STATE 
			 PROBLEM IN A WILSON LOOP CONTEXT}

\author{ N. BRAMBILLA and G. M. PROSPERI}

\address{Dipartimento di Fisica dell' Universit\`a di Milano and
I.N.F.N.\\
Via Celoria 16, 20133 Milano}

\maketitle\abstracts{ The paper is on the line of the path integral
technique developped in preceding articles. Taking advantage of 
a semirelativistic and  a full relativistic representation 
of the quark propagator in an external field we present  an unified derivation
of the semirelativistic potential and of a Bethe--Salpeter like equation
for the quark--antiquark system. We consider three different  models 
for the evaluation  of the Wilson loop: the Modified Area Law  model 
(MAL), the Stochastic Vacuum Model (SVM)  and the Dual QCD (DQCD).
We compare  the corresponding potentials  and show that they  all 
agree at the short and the long distances.
 In the case of the Bethe--Salpeter equation
 we treat explicitly only  the MAL model and give an explicit  expression 
 for the kernel. Then we show that an effective mass operator  can be obtained 
which agrees with the MAL potential  in the semirelativistic limit.
In the light quark mass limit  this mass operator produces straight Regge 
trajectories with Nambu--Goto slope in agreement with the data.
Finally we briefly discuss the mass independence of the hyperfine
splitting in the heavy--light case.}

\section{Introduction}
 
\indent
In preceding papers \cite{bp}, \cite{co94} we  have shown 
how, using 
certain path integral representations of the 
quark-antiquark gauge invariant Green function $ G^{\rm gi} (x_1,x_2; y_1,y_2) 
$ is possible to derive a semirelativistic potential for heavy quarks or a 
Bethe-Salpeter like (BS) equation for light and heavy quarks in terms of the Wilson
loop integral $W$ and its functional derivatives.
Such representations are the consequence of corresponding representations
for the quark propagator in an external gauge field.

   We have used two different path integral representations for the propagator.
The first one is expressed in terms of a  ${1\over m}$ expansion
 and quark paths in ordinary phase 
space\cite{bp}. This is appropriate for the 
derivation of the heavy quark potential. The second one is expressed 
in terms of covariant
space-time quark paths \cite{peskin}, \cite{co94}, \cite{Simonov}
and it is useful for full relativistic developments.

   Since up to now it is not possible to evaluate  the Wilson loop analytically 
from first principles, to obtain explicit expressions one has to rely on models
and/or lattice simulations. In this paper we shall discuss three different models: the 
Modified Area Law model (MAL), in which $i \ln W $ is written as the sum of a 
perturbative term, an area term and a perimeter term \cite{bp}, \cite{co94};
the Stochastic Vacuum Model (SVM) \cite{Dosch}, which is based on a cumulant
expansion; the dual QCD (DQCD) \cite{bbz}, which introduces effective fields and
an effective lagrangian which should simulate what is believed to be the real
mechanism for  the formation of the 
confining flux tube. For each of these models a different 
potential can be obtained. We shall  
compare such potentials  and show that they 
agree for large and small distances, but  differ in the   
intermediate range. 
   Up today 
    an explicit BS equation can  be obtained only  in the MAL case. 
    Such equation is written  in terms of  a second
order Green function $ H(x_1,x_2;y_1,y_2) $ to which 
 $ G^{\rm gi} (x_1,x_2; y_1,y_2) $ is related 
 for large $x_j^0-y_j^0 (j=1,2)$ simply by the application 
 of Dirac type differential operator. The BS kernel
$ I(x_1,y_1; x_2,y_2) $ is expressed in principle 
as an expansion in the strong coupling 
constant $\alpha_{\rm s}$ and the string tension $\sigma$ (more precisely
the quantity $\sigma a^2$, $a$ being the typical radius of a bound state).
In principle the method could be applied also to SVM and DQCD models but the 
actual calculations would be very complicated and have not been performed.
In practice even the MAL kernel has been worked out only at the lowest order.
We show that starting from
    the MAL BS equation a squared  ($M^2$) or a linear
  ($M$)   relativistic effective  mass operator involving 
  certain relativistic potential $U$ or $V$
can be obtained in the instantaneous limit. Spectrum 
calculation using such expressions are in progress. Obviously, in the 
semirelativistic limit $V$ reproduces the MAL potential.
On the contrary, if one 
retains  full relativistic kinematic but 
neglects the spin dependent terms, $M$ 
coincides with the hamiltonian of the Relativistic Flux Fube Model, discussed
in Ref. \cite{olsson}, up to ordering prescription and in the limit of 
vanishing quark masses gives Regge trajectories with slope $\alpha^\prime=
{1 \over 2 \pi \sigma}$ as in Nambu-Goto string  which are  in agreement with 
phenomenology ($ \sigma \sim 0.18 {\rm GeV}^2$, 
$\alpha^\prime \sim 0.88$). 
Furthermore, if the spin dependent terms are kept but the limit of 
heavy--light quarks
is considered, spin symmetries similar to those discussed by Kaidalov 
\cite{kaidalov} are obtained.
   In Sec. 2 we 
   give the basic equation for the definition of the potential and BS 
   equation; in Secs. 3 and 4 we review the three mentioned models for the 
evaluation of the Wilson loop and compare the form of the 
corresponding potentials; in Sect. 5 we 
report the  BS kernel 
 in the MAL case and discuss the mass operator.
The present paper is mainly based on \cite{bbbpz}, \cite{Dosch},
\cite{bv}, \cite{bsnoi} but it is in part original  in the 
presentation and contains some new results.

\section{    Definition of the semirelativistic potential and BS equation}

   In the  usual functional integral formulation, after integration on the 
fermionic variables, the gauge invariant quark-antiquark 
Green function can be written
\begin{equation}
 G^{\rm gi}(x_1,x_2;y_1,y_2) =
 \frac{1}{3} {\rm Tr} \langle U(x_2,x_1)
 S^{(1)}(x_1,y_1|A) U(y_1,y_2) C^{-1}
S^{(2)} (y_2,x_2|A) C \rangle \> .
\label{eq:ventitre}
\end{equation}
 Here $C$ is  the charge coniugation matrix, $ U(a,b) $ denotes the Schwinger string
\begin{equation}
U(b,a) = {\rm P} \exp \big \{ ig \int_a^b dx^\mu A_\mu (x) \big \}
\end{equation}
(with $ A_\mu (x) = {1\over 2} \lambda^a A_\mu ^a (x) $ and ${\rm P}$ the 
path ordering
operator over the color matrices), $ S(x,y;A) $ is the quark
propagator in an external field
\begin{equation}
( i \gamma^\mu D_\mu - m) S(x,y;A) = \delta^4(x-y)
\label{dirac}
\end{equation}
( $ D_\mu = \partial_\mu - igA_\mu (x) $) and the notation
\begin{equation}
\langle f[A] \rangle = \frac{\int {\cal D}A\, M_f(A)
f[A] e^{iS_{\rm YM}[A]}}
{\int {\cal D}A\, M_f(A) e^{iS_{\rm YM}[A]}} \> ,
\label{eq:med}
\end{equation}
is used, $ M_f (A) $ being the fermionic determinant 
and $S_{\rm YM}[A]$ the pure Yang--Mills action.

For a closed curve $\Gamma$     
we set across the paper (Wilson loop correlator)
    $$ W(\Gamma) = {1 \over 3} \langle {\rm Tr}\, {\rm P}\exp \{ \oint_\Gamma
	    dx^\mu A_\mu (x) \} \rangle \, . 
	    \label{defwil}$$
We are interested in loops formed by 
a line $\Gamma_1$ joining $y_1$ to $x_1$ (the quark path),
 another line $\Gamma_2$ 
joining $x_2$ to $y_2$ (the reverse antiquark path) and 
 two straight lines  connecting $x_1$ with $x_2$
and  $y_2$ with  $y_1$.
In term of these 
  the semirelativistic quark--antiquark potential is defined by\cite{bp},
\cite{co94}
\begin{eqnarray}
\int_{y_0}^{x_0} dt V_{{\rm Q} \bar{{\rm Q}}} &=& 
i \log  W(\Gamma)  
- \sum_{j=1}^2 \frac{g}{2 m_j} \int_{{\Gamma}_j}dx^{\mu} 
\left( \sigma_j^l \, 
\langle\!\langle \hat{F}_{l\mu}(x) \rangle\!\rangle  
-\frac{1}{2m_j} \sigma_j^l \varepsilon^{lkr} p_j^k \, 
\langle\!\langle F_{\mu r}(x) \rangle\!\rangle \right.
\nonumber\\
&~& \quad - \left. \frac{1}{8m_j} \, 
\langle\!\langle D^{\nu} F_{\nu\mu}(x) \rangle\!\rangle  \right)
- \frac{1}{2} \sum_{j,j^{\prime}=1}^2 \frac{ig^2}{4m_jm_{j^{\prime}}}
{\rm T_s} \int_{{\Gamma}_j} dx^{\mu} \, \int_{{\Gamma}_{j^{\prime}}} 
dx^{\prime\sigma} \, \sigma_j^l \, \sigma_{j^{\prime}}^k
\nonumber\\
&~& \quad \times \left( \, 
\langle\!\langle \hat{F}_{l \mu}(x) \hat{F}_{k \sigma}(x^{\prime})
\rangle\!\rangle  - \, 
\langle\!\langle \hat{F}_{l \mu}(x) \rangle\!\rangle
\, \langle\!\langle \hat{F}_{k \sigma}(x^{\prime}) \rangle\!\rangle  \right) 
\> ,
\label{potential}
\end{eqnarray}
with $y_1^0=y_2^0=y^0, x_1^0=x_2^0=x^0$, ${\rm T}_s$ time ordering 
prescription  on the spin matrices, 
$\hat F_{\mu \nu}={1 \over 2} \epsilon_{\mu \nu \rho \sigma} 
F^{\rho \sigma}$ and
\begin{equation}
\langle\!\langle f(A) \rangle\!\rangle \equiv
{\int {\cal D} A \, e^{iS_{\rm YM} (A)} 
{\rm Tr \>}{\rm P \>} \{ f(A) 
\exp \left[i g \oint_{\Gamma} dx^\mu A_\mu (x) \right] \}
\over \int {\cal D} A\,  e^{iS_{\rm YM} (A)} {\rm Tr \>}{\rm P \>}
{\exp \left[i g \oint_{\Gamma} dx^\mu A_\mu (x) \right] } } \>.
\label{doppio}
\end{equation}
 Eq.~\ref{potential} is obtained solving  Eq.\ref{dirac} for
 $x_0 >y_0$ by a Foldy--Wouthysen transformation and  path integral 
technique and comparing the resulting expression for $G^{gi}$ with
the two particle Schr\"odinger equation path representation.
Notice that if 
   $\delta S^{\mu \nu} (z)= {1\over 2}(dz^\mu \delta z^\nu 
 - dz^\nu \delta z^\mu) $ denotes the surface swept by the quark path as a 
consequence of an infinitesimal variation $z(\tau) \rightarrow z(\tau) +
 \delta z(\tau)$ one has
	$$ \delta \Big \{{\rm P} 
	\exp [ig \int_y^x dz^\mu A_\mu (z) ] \Big \} =
     ig  {\rm P}\int\delta S^{\mu \nu} (z) F_{\mu \nu} (z)  
     \exp \{ig \int_y^x dz^{\mu \prime} A_\mu (z^\prime) \} . $$ 
From this  it follows immediately
\begin{eqnarray}
& &  g \langle\!\langle F_{\mu\nu}(z_j) \rangle\!\rangle = (-1)^{j+1}
{\delta i \log \langle W(\Gamma) \rangle \over \delta S^{\mu\nu} (z_j)} \> ,
\quad \quad (j=1,2)
\label{e20}
\\
& & g^2 \left(\langle\!\langle F_{\mu\nu}(z_1) 
F_{\lambda\rho}(z_2) \rangle\!\rangle 
- \langle\!\langle F_{\mu\nu}(z_1) \rangle\!\rangle 
  \langle\!\langle F_{\lambda\rho}(z_2) \rangle\!\rangle \right)
= - i g {\delta\over \delta S^{\lambda\rho}(z_2)} 
\langle\!\langle F_{\mu\nu}(z_1) \rangle\!\rangle.\nonumber\\
  & & 
\label{e21}
\end{eqnarray}
etc. and so everything in the right hand side of Eq.~\ref{potential} 
can be expressed
in terms of $W$. Notice, however, that, both through the expression of 
$ i \ln W$
(e.g. cf.~\ref{pertw}) and the spin dependent terms, the right 
hand side of~\ref{potential} contains integrals on two separate times $t_1$ and
$t_2$, while, by assumption, the left hand side contains one single time $t$.
One can dispose of the above situation by integrating explicitely the relative
time $\tau = t_1 - t_2$ after setting
\begin{equation}
{\bf z}_j(t_j)  =  {\bf z}_j(t) + {1\over 2} (-1)^{j+1}  \tau 
\dot{\bf z}_j(t) + {1\over 8}
\tau^2 \ddot{\bf z}_j(t) + \dots 
\end{equation}
keeping only the appropriate order term and eliminating the second time 
derivatives by partial integration (notice that $\dot{z}_j^0 =1;
\ddot{\bf z}_j = O(\dot{\bf z}_j^2)$).

On the contrary,     
if one sets
\begin{equation}
S(x,y;A) = (i\gamma^\mu D_\mu +m) \Delta (x,y;A)\,   ,
\label{reprdirac}
\end{equation}
Eq. ~\ref{dirac} gives
\begin{equation}
(D_\mu D^\mu +m^2 - {1\over 2} \sigma^{\mu\nu} F_{\mu\nu} )\Delta(x,y;A)
= - \delta^4(x-y)\,  .
\label{reprsec}
\end{equation}
Then, for large time intervals  one can write
 \begin{equation}
  G^{gi}(x_1,x_2;y_1,y_2)= 
 (i\gamma_1^\mu {\partial}_{1\mu} +m_1) 
(i\gamma_2^\nu {\partial}_{2\nu} +m_2) H(x_1x_2;y_1 y_2),
\label{reldue}
\end{equation} 
with 
\begin{equation}
 H(x_1 x_2; y_1 y_2) = 
 {1\over 3}{\rm Tr} \langle U(x_2, x_1) \Delta^{(1)}(x_1,y_1;A)
 U(y_1, y_2) \tilde{\Delta}^{(2)}(y_2,x_2;\tilde{A})\rangle  
 \label{hqua}
 \end{equation}
where the  tilde denotes transposition of the color matrices.
 Furthermore, let us assume that $ i\ln W(\Gamma)$ can be set in 
 the general form
\begin{equation}        
 i \ln W = {\lambda \over 2} \sum_{i,j} \int_0^{s_i} d \tau_i 
	   \int_0^{s_j} d \tau _j ^\prime E(z_i,z_j ^\prime,\dot z_i, 
	   \dot z_j ^\prime) \quad + \quad \dots \quad ,   \label{rapgen}
\end{equation}
where the dots stand for $\lambda ^2$ 
terms involving similar fourth order
integrals, $\lambda^3$ terms involving six order integrals and so on 
($\lambda$ being a parameter that has been introduced for convenience)
and $ E$ is an homogenous  function of degree $0$ in $\dot{z}_1$,
$\dot{z}_2$. Then, using for $\Delta(x,y;A)$ the covariant path integral 
Feynman--Schwinger representation \cite{peskin}, \cite{Simonov},
\cite{co94}, one can obtain the inhomogeneous BS equation\cite{bsnoi}
\begin{eqnarray}
& & 
H(x_1,\,x_2;\,y_1,\,y_2) =  
H^{(1)}(x_1-y_1)\, H^{(2)}(x_2-y_2)\, - 
\,i \int d^4\xi_1 d^4\xi_2 d^4\eta_1 d^4\eta_2 
\nonumber \\
& & 
\quad  H^{(1)}(x_1-\xi_1)\, 
H^{(2)}(x_2-\xi_2)\, I(\xi_1,\,\xi_2;\,\eta_1,\,\eta_2) \, 
H(\eta_1,\,\eta_2;\, y_1,\, y_2)   
\label{bsh}
\end{eqnarray}
with the kernel $I$ expressed as expansion in $\lambda$
and $H(x-y)=\langle 
\Delta(x,y;A)\rangle$.
At the first order in $\lambda$ one has 
\begin{equation}
I(\xi_1, \xi_2 , \eta_1, \eta_2 ) = -4  \int { d^4 k_1 d^4 k_2 \over (2 
\pi )^8}  R( {\xi_1 + \eta_1 \over 2}, { \xi_2 +\eta_2 \over 2}, 
k_1 , k_2 ) \exp \big  \{-i [ (\xi_1 -\eta_1 )
 k_1 + (\xi_2 -\eta_2 ) k_2 ]\big \},
\label{eq:kernr}
\end{equation}
with $R$ defined by 
\begin{equation}
{\cal S}_0^{s_1} {\cal S}_0^{s_2} \int d\tau_1 \int d\tau_2 E(z_1,
z_2, p_1,p_2) ({\cal S}_0^{s_1} {\cal S}_0^{s_2})^{-1} =
\int_0^{s_1} d\tau_1 \int_0^{s_2} d\tau_2 R(z_1,z_2,p_1,p_2)
\end{equation}
  having set 
 $
 {\cal S}_0^s = {\rm T}_s \exp \big [ - {1\over 4} \int_0^s 
 d\tau \sigma^{\mu\nu} {\delta \over \delta S^{\mu \nu}(z)} \big ]
$.

\section{          Models for Wilson Loop Evaluation}

 \subsection{Modified Area Law Model}\label{mal}

   As mentioned, MAL model consists 
   in assuming $ i \ln W $ to be the sum of its
perturbative value,  area and  perimeter term
\begin{equation}
i \log \langle W (\Gamma) \rangle =  
i\log \langle W (\Gamma) \rangle_{\rm pert} +\sigma S_{\rm min} 
+{1\over 2} C P  \ ,
\label{mod1}
\end{equation}
where obviously $S_{\rm min}$ denotes the minimal area enclosed by the loop 
$\Gamma$ and $P$ its length as suggested by the Wilson strong 
coupling limit \cite{lsd},
\cite{co94}, while $\sigma$ and $C$ are in practice treated as independent 
adjustable parameters (which must however agree with  the lattice 
simulations).
   At lowest order one has
\begin{equation}
i (\ln W)_{\rm pert} = {4\over 3} g^2 \int_{y_{10}}^{x_{10}} dt_1
\int_{y_{20}}^{x_{20}} dt_2 \dot{z}_1^\mu \dot{z}_2^\nu
\,  D_{\mu\nu}(z_1-z_2)
\> .
\label{pertw}
\end{equation}
Furthermore
\begin{equation}
S_{\rm min}  =  \min \int_{t_{\rm i}}^{t_{\rm f}}
dt \, \int_0^1  d\alpha
 \, \left[-
\left( \frac{\partial u^{\mu}}{\partial t} \frac{\partial u_{\mu}}
{\partial t} \right) \left( \frac{\partial u^{\mu}}{\partial \alpha}
\frac{\partial u_{\mu}}{\partial \alpha} \right) + \left(
\frac{\partial u^{\mu}}{\partial t} \frac{\partial u_{\mu}}
{\partial \alpha} \right)^2 \right]^{\frac{1}{2}} \ ,
\label{min1}
\end{equation}
\begin{equation}
P= \sum_{j} \int_{y_{j0}}^{x_{jo}} dt_j [\dot{z}_j^\mu 
\dot{z}_{j\mu}]^{1\over 2} \ ,
\label{per1}
\end{equation}
where  $u^\mu= u^\mu (t, \alpha)$ with $y^0 < t < x^0,\; 
0<\alpha <1$ 
is the equation  of an arbitrary surface enclosed by
 $\Gamma$ and satisfying therefore
boundary conditions
\begin{equation}
u^\mu(t,0)=z_2^\mu(t) \ , 
\quad \quad \quad u^\mu(t,1)=z_1^\mu(t).
\end{equation}
In practice an equal time straight line approximation is commonly adopted
for $S_{\rm min}$. This amounts to setting
\begin{equation}
u^0(t,\alpha)=t  \ ,  \quad \quad \quad
{\bf u}(t,\alpha) = \alpha~{\bf z}_1(t) + (1-\alpha)~ {\bf z}_2(t) \>,
\label{straight}
\end{equation}
\begin{eqnarray}
& & \quad \quad \quad S_{\min} =
\int_{t_{\rm i}}^{t_{\rm f}} dt \,  r \int_0^1 d\alpha \, 
[1-(\alpha~\dot{{\bf z}}_{1 \rm T} + (1-\alpha)~
 \dot{{\bf z}}_{2 \rm T} )^2]^{\frac{1}{2}} = \\
&  &=\int_0^{s_1} d\tau_1 \int_0^{s_2} d\tau_2 \delta(z_1^0 - z_2^0) 
 \vert {\bf z}_1 -{\bf z}_2 \vert \int_0^1 \big \{ \dot{z}_1^0 \dot{z}_2^0 
 - [\alpha \dot{\bf z}_{1{\rm T}} \dot {z}_2^0 +(1-\alpha )
 \dot{\bf z}_{2 {\rm T}} \dot{z}_1^0 ]^2 \big \}^{1\over 2}\nonumber
 \label{min2}
 \end{eqnarray}
 with $ z^h_{j{\rm T}}= (\delta^{hk} - \hat{r}^h \hat{r}^k ) z_j^k$
 (obviously in the first step $\dot{z}_j={d z_j\over dt}$,
 in the second one $\dot{z}_j^\mu={dz_j^\mu\over d\tau_j}$).
This equation it is exact to order $1/m^2$ and 
for particular geometries. The second step is given in terms of a 
covariant parametrization of the quark and the antiquark paths and is useful
in the derivation of the BS equation; however it is not Lorentz 
invariant and it  is  assumed to be true in the center of mass 
frame.

\subsection{ Stochastic Vacuum Model }\label{svm}

    Using the non abelian Stokes theorem  and cumulant 
expansions, one can write 
\begin{eqnarray}
\langle W(\Gamma) \rangle &=& 
\left\langle {\rm P} \>
\exp \left( ig \int_S dS^{\mu\nu}(u) F_{\mu\nu}(u,x_0) \right) \right\rangle 
\label{stokes}\\ 
&=& 
 \exp \sum_{j=1}^{\infty} {(ig)^j\over j!} 
\int_S dS^{\mu_1\nu_1}(u_1) \dots 
\nonumber\\
&~& \quad\quad\quad 
\int_S dS^{\mu_j\nu_j} (u_j) 
\langle F_{\mu_1\nu_1}(u_1,x_0) \dots F_{\mu_j\nu_j}(u_j,x_0)
\rangle_{\rm cum}  \>.
\label{cluster}
\end{eqnarray}
where $F_{\mu \nu}(u,x_0)= U(x_0,u) F_{\mu\nu} U(u,x_0)$ and 
 the cumulants $\langle \dots \rangle _{\rm cum}$ are defined by
\begin{equation}
\langle F(1)\rangle_{\rm cum} = \langle F(1) \rangle \>, \quad\quad 
\langle F(1) F(2) \rangle_{\rm cum} = 
\langle F(1) F(2)\rangle  - \langle F(1) \rangle \langle F(2)\rangle
\>, ~~ \dots 
\nonumber
\end{equation}
 $S$ is an arbitrary surface enclosed by $\Gamma$, $x_0$ a reference 
point on $S$, ${\rm P}_S$ an ordering prescription 
on S and the $U(u,x_0)$ are the Schwinger strings
\cite{Dosch}.

   The basic approximation consists in assuming that the second cumulant is 
dominant and actually independent of $x_0$. Then, since the first cumulant 
vanishes trivially, one can write
\begin{equation}
\log \langle W(\Gamma) \rangle  = -{g^2 \over 2} 
\int_S dS^{\mu\nu}(u) \int_S dS^{\lambda\rho} (v)  
\langle F_{\mu\nu}(u,x_0)  F_{\lambda\rho}(v,x_0) \rangle_{\rm cum}
\> .
\label{svm}
\end{equation}
and, taking into account Lorentz invariance,
\begin{eqnarray}
& & g^2 \langle F_{\mu\nu}(u,x_0)  F_{\lambda\rho}(v,x_0) \rangle_{\rm cum}
= g^2\langle F_{\mu\nu}(u,x_0) F_{\lambda\rho}(v,x_0) \rangle \> 
\nonumber\\
&& = {\beta} \Bigg\{
(\delta_{\mu\lambda}\delta_{\nu\rho} - 
\delta_{\mu\rho}\delta_{\nu\lambda})D ( (u-v)^2 ) 
+ {1\over 2}[
{\partial\over\partial u_\mu} ( (u-v)_\lambda\delta_{\nu\rho} 
- (u-v)_\rho\delta_{\nu\lambda} )+\nonumber \\ 
& & + {\partial\over\partial u_\nu} ( (u-v)_\rho\delta_{\mu\lambda} 
- (u-v)_\lambda\delta_{\mu\rho} ) ] D_1 ( (u-v)^2 )
\Bigg\}
\label{cum2} 
\end{eqnarray}
where $
\beta\equiv {g^2\over 36} 
{\langle {\rm Tr \>} F_{\mu\nu}(0) F_{\mu\nu}(0) 
\rangle \over D(0) + D_1(0)}$
and  $D$ and $D_1$ are unknown  functions.
    In the euclidean space, at the lowest order in perturbation theory one finds
	$$ D^{\rm pert}(x^2) = 0 \quad \quad D_1^{\rm pert}(x^2) = 
	      {16 \alpha_{\rm s} \over 3 \pi}{1 \over x^4} \ ,  $$
which is supposed to give the behaviour of the functions for 
$x \rightarrow 0$.  A
good parametrization for the long range region seems to be
\begin{eqnarray}
\beta~D^{\rm LR}(x^2) &=& d~e^{-\delta|x|} \>,
\quad \> \delta = (1 \pm 0.1) ~{\rm GeV} \>, 
\quad d = 0.073 ~{\rm GeV}^4 \>,
\label{Dlr}\\
\beta~D_1^{\rm LR}(x^2) &=& d_1~e^{-\delta_1|x|} \>,
\quad \delta_1 = (1 \pm 0.1) ~{\rm GeV} \>, 
\quad d_1 = 0.0254 ~{\rm GeV}^4 \>,
\nonumber
\end{eqnarray}
where the value of gluonic condensate and lattice simulations have been combined
\cite{Digiacomo}.

\subsection{Dual QCD}\label{dqcd}

   Dual QCD \cite{bbz} is an effective theory described by a lagrangian density 
${\cal L}_{\rm eff}$ in which the fundamental variables are  an octet of dual
potential $C^\mu$ coupled to a classical quark source and
 to three octets of scalar fields $B_i$ carring 
"magnetic" color charge. The "monopole" fields
$B_i$ develop  nonvanishing 
vacuum expectation values $B_{0i}$ that give rise to a dual Meissner effect and 
provide a concrete realization of the Mandelstam-t'Hooft picture of confinement.
We set 
\begin{equation}
  W_{\rm eff} (\Gamma) =
   {
   \int {\cal D} C_\mu {\cal D} \phi {\cal D} B_3
    e ^ {i \int dx [ {\cal L}_{eff} (G_{\mu\nu}^S) + {\cal L}_{GF} ] }
  \over
  \int {\cal D} C_\mu {\cal D} \phi {\cal D} B_3
    e ^ {i \int dx [ {\cal L}_{eff} (G_{\mu\nu}^S=0) + {\cal L}_{GF} ] }
   },
  \label{weff}
\end{equation} 
where
 \begin{equation} {\cal L}_{\rm eff} = 2 {\rm Tr} 
\left[ - {1\over 4} {\bf G}^{\mu\nu} {\bf G}_{\mu\nu}
+ {1\over 2} ({\cal {\cal D}}_\mu {\bf B}_i)^2\right] - W({\bf B}_i),
\label{leff}
\end{equation}
\begin{equation}
{\cal D}_\mu {\bf B}_i = \partial_\mu {\bf B}_i - i g_M [ {\bf C}_\mu, {\bf
B}_i],
\end{equation}
\begin{equation}
{\bf G}_{\mu\nu} = \partial_\mu {\bf C}_\nu - \partial_\nu {\bf C}_\mu - i g_M
[{\bf C}_\mu, {\bf C}_\nu] + {\bf G}_{\mu\nu}^{S},
\label{dft}
\end{equation}
${\cal L}_{GF}$ is the gauge fixing term,
 $g_M={2\pi \over g}$ is the magnetic coupling constant and
\begin{equation}
{\bf G}_{\mu \nu}^{S}(x) = g {\lambda_8 \over \sqrt{3}} 
\varepsilon_{\mu\nu\rho\sigma} \int_0^1 d\alpha {\partial u^\rho \over
\partial \alpha} {\partial u^\sigma \over \partial t} \delta(x-u(\alpha, t)),
\label{string}
\end{equation}
$ u^\mu=u^\mu(t,\alpha) $
 being the surface swept by the Dirac string connecting the
quark and the antiquark. The Higgs potential $W({\bf B}_i)$ (which we do not 
give explicitly) has a minimum at non zero values $B_{0i}$ of the form
\begin{equation}
{\bf B}_1= B_0 \lambda_7 \> ,
\quad \quad {\bf B}_2 = - B_0 \lambda_5 \> , \quad \quad
{\bf B}_3 = B_0 \lambda_2\> .
\label{mind}
\end{equation}

   In our present context the basic assumption\cite{bbbpz}
 is that for large $\Gamma$ 
\begin{equation}
W(\Gamma) \simeq W_{\rm eff}\  .
\label{dueventi}
\end{equation}

    To calculate explicitly $W_{\rm eff}$ even for restrict class of quantum
fluctuation is not a trivial task. However, in the so called classical 
approximation this quantity is given by the expression 
$ \exp \{ i \int
d^4x  {\cal L}_{\rm GF}(G_{\mu \nu}^S) \}$ evaluated
for a classical solution of the field equations of the form
\begin{eqnarray}
& & \quad \quad \quad 
\bar{\bf C}^\mu   =   \bar{C}^\mu {\lambda_8\over \sqrt{3}}\nonumber\\
& & \bar{\bf B}_1 =\bar{B} \lambda_7\quad \quad \quad \bar{\bf B}_2 = - \bar{B}
\lambda_5 \quad \quad  \quad \bar{\bf B}_3 = B^\prime \lambda_2 
\end{eqnarray}
where $\bar{C}^\mu$, $\bar{B}$ and $\bar{B}^\prime$ are
 appropriate solutions of 
the equations
\begin{eqnarray}
\partial^\mu (\partial_\mu \bar{C}_\nu - \partial_\nu \bar{C}_\mu)& =& -
\partial^\mu G_{\mu \nu}^S - 6 g^2 \bar{C}_\mu \bar{B}^2
\nonumber\\
(\partial_\mu +i g \bar{C}_\mu)^2 \bar{B}&  =& 
- {1\over 4} {\delta U\over\delta\bar{B}} \quad \quad \quad 
\partial^2 {B}^\prime = - {1\over 4} {\delta U\over\delta B^\prime}
\label{motoeq} 
\end{eqnarray}
satisfying the asymptotic conditions $\bar{C}^\mu \rightarrow 0,\ \bar{B},
 \bar{B}^\prime \rightarrow B_0 $. Then Eqs.~\ref{motoeq} 
can be solved numerically  and analytically parametrized
\cite{bbz}.

\section{      Semirelavistic Potential}

    On the basis of general invariance principle the quark-antiquark potential
up to order $1/m^2$ can be written as
\begin{equation} 
V_{{\rm Q} \bar {\rm Q}} = V_0(r) + V_{\rm VD}({\bf r}, {\bf p}_1,
{\bf p}_2) + V_{\rm SD}({\bf r}, {\bf p}_1, {\bf p}_2, {\bf \sigma}_1, 
{\bf \sigma}_2)
\label{defpot}
\end{equation}
with
\begin{eqnarray}
V_{\rm VD}({\bf r}(t)) &=&  {1\over m_1 m_2} 
\left\{ {\bf p}_1\cdot{\bf p}_2 V_{\rm b}(r) 
+ \left( {1\over 3} {\bf p}_1\cdot{\bf p}_2 - 
{{\bf p}_1\cdot {\bf r} \>~ {\bf p}_2 \cdot {\bf r} \over r^2}\right) 
V_{\rm c}(r) \right\}_{\rm Weyl} 
\nonumber \\
&+& \sum_{j=1}^2 {1\over m_j^2}
\left\{ p^2_j V_{\rm d}(r) 
+ \left( {1\over 3} p^2_j - 
{{\bf p}_j\cdot {\bf r} \>~ {\bf p}_j \cdot {\bf r} \over r^2}\right) 
V_{\rm e}(r) \right\}_{\rm Weyl} \, , 
\label{vd}
\end{eqnarray}
and
\begin{eqnarray}
V_{\rm SD} &=& 
{1\over 8} \left( {1\over m_1^2} + {1\over m_2^2} \right) 
\Delta \left[ V_0(r) +V_{\rm a}(r) \right] 
\nonumber\\
&+& \left( {1\over 4 m_1^2} {\bf L}_1 \cdot {\bf \sigma}_1 
     - {1\over 4 m_2^2} {\bf L}_2 \cdot {\bf \sigma}_2 \right) 
       {1\over r}  {d \over dr} \left[ V_0(r)+ 2 V_1(r) \right]
\nonumber \\
&+&
{1\over 2 m_1 m_2}
\left( {\bf L}_1 \cdot {\bf \sigma}_2 - {\bf L}_2 \cdot {\bf \sigma}_1 
\right) 
{1\over r} {d \over dr} V_2(r) 
+{1\over 4 m_1 m_2}  
\left( { {\bf \sigma}_1\cdot{\bf r} \> {\bf \sigma}_2\cdot{\bf r}\over r^2} 
- {1\over 3} {\bf \sigma}_1 \cdot {\bf \sigma}_2 \right) V_3(r) 
\nonumber \\
&+& {1\over 12 m_1 m_2} 
{\bf \sigma}_1 \cdot {\bf \sigma}_2 \, V_4(r) \>,
\label{sd}
\end{eqnarray}
in terms of functions of $r$ alone.

   The various functions $V_0(r), \dots V_e(r)$ can be  
   written in terms of momenta of fields \cite{lsd}, \cite{bp}. Such expressions
are very useful for lattice simulations \cite{lattice}.

   Finally, one can work out a  potential in  explicit 
form for any of the models considered in Sec. 3.
In Tab. 1 for the MAL model we report the complete expression, 
while for SVM and DQCD we report only the long range 
behaviour (we omit exponentially  vanishing terms). In terms of the 
parametrization  of Eq. 30 the constants occurring in the second 
column are given by \cite{bv}

\begin{table}[t]
\caption{\label{tab:exp} Complete MAL potential and long 
distance SVM and DQCD potentials }
\vspace{0.4cm}
\begin{center}
\footnotesize
\begin{tabular}{|c|c|c|l|}
\hline
& & & \\
&
MAL & SVM & DQCD
\\ \hline
$V_0$ &
$-{4\over 3} {\alpha_s\over r}+ 
\sigma r + C $ & $ \sigma_2 r + {1\over 2} C_2^{(1)} - C_2 $&
$\sigma r -0.646 \sqrt{\sigma \alpha_s}$
\\ & & & \\
$\Delta V_a $ & 0 & Self-energy terms &  
\\ & &  & \\
${V_1^\prime}$ & $\sigma $ & $-\sigma_2 + {C_2\over r}$ &
$-\sigma + {0.681 \over r} \sqrt{\sigma \alpha_s}$\\
${ V_2^\prime}$ & 
${4\over 3}
{\alpha_s\over r^2}$ & ${C_2 \over r}$ & $ {0.681\over r} \sqrt{\sigma
\alpha_s}$ \\
& & & \\
$V_3$ & $ 4 {\alpha_s\over r^3}$ & 0 & 0\\ 
& & & \\
$ V_4$ & $ 32 \pi \alpha_s \delta^3({\bf r})$& 0& 0 \\
& & & \\
$V_b$ & ${8\over 9} {\alpha_s\over r} 
- {1\over 9} \sigma r$ & $- {1\over 9} \sigma_2 r - {2\over 3}
{ D_2\over r} + {8\over 3} {E_2\over r^2}$ & $ -0.097 \sigma r -0.226 
\sqrt{\sigma \alpha_s}$ \\
& & & \\
$V_c$ & $ -{2\over 3}{\alpha_s\over r}
- {1\over 6} \sigma r$ & $ -{1\over 6}\sigma_2 r - {D_2\over r}
+ {2\over 3} {E_2\over r^2} $& $ -0.146 \sigma r -0.516 \sqrt{\sigma \alpha_s}$
\\
& & & \\
$V_d$ & $ -{1\over 9} \sigma r - {1\over 4} C$ & $ - {1\over 9} \sigma_2 r +
 {1\over 4} C_2 - {1\over 8} C_2^{(1)} + {1\over 3} {D_2 \over r} - {2\over 9}
{E_2\over r^2} $ & $ -0.118 \sigma r +0.275 \sqrt{\sigma\alpha_s}$\\
$V_e$ & $ -{1\over 6} \sigma r$ &$ -{1\over 6} \sigma_2 r + {1\over 2}
{D_2 \over r} - {1\over 3} {E_2\over r^2}$ & $- 0.177 \sigma r + 0.258
 \sqrt{\sigma \alpha_s} $\\
& & &  \\ \hline
\end{tabular}
\end{center}
\end{table}
\begin{eqnarray}
\sigma_2= {\pi d \over \delta^2}    \quad && 
\quad C_2= {4 d \over \delta^3}   \\
C_2^{(1)}= { 4 d_1 \over \delta_1^2 }  \quad
& & \quad D_2={3\pi d \over \delta^4}  \quad E_2= {32 d\over 
\delta^5} \nonumber
\label{svmpar}
\end{eqnarray}
Complete expressions even for SVM and DQCD can be found in \cite{Simonov}
\cite{bv} and \cite{bbz}. Notice that the short range behaviour 
must agree for the three potentials by construction. 
In the long range region, as one can see,
  there is essentially agreement among 
  the leading terms. Some minor discrepancies 
  in the DQCD case are possibly due to numerical 
inaccuracy. On the contrary the discrepancies in the subleading terms seem to 
be more significant. Notice however the complete similarities between the
SVM and the DQCD columns for the first six lines of the table.
Obviously the three models definitely differ at intermediate ranges.

   For what concerns the numerical values of the constants one may recall 
that from heavy quarkonia fitting typical values are $\sigma
\simeq 0.18 {\rm GeV} ^2$ and $\alpha _{\rm s}\simeq 0.37$. On the other
 side
if from lattice simulation 
 we assume  [cf. Eq.30] 
the marginal values
 $\delta=\delta_1=1.1$, we find  $\sigma \simeq 0.19 {\rm GeV} ^2, \quad C_2
\simeq 0.22 {\rm GeV}, \quad C_2^{(1)} \simeq 0.08 {\rm GeV}, \quad D_2 
\simeq 0.47, \quad E_2 \simeq 1.45 {\rm GeV}^{-1}$. 
In this case the agreement between SVM  and  DQCD 
 is therefore remarkable  for the first six lines  even from the 
numerical point of view. Finally, notice that in the MAL case the spectrum is very
little sensible to the values of the constant $C$. In fact due to the combined
form of $V_0$ and $V_D$ a variation of $C$ can be reabsorbed in refinition of
the quark masses $m \rightarrow m + {\delta C \over 2}$. This is not true in the
SVM and DQCD cases due to the form of $V_1$ and $V_2$.

\section{    Bethe-Salpeter kernel and effective mass operator}

    Let us begin to notice that in the MAL case, neglecting the perimeter 
term and taking into account Eqs.~\ref{pertw} and 24, Eq. 19 
can turn out to be of the general form of Eq.~\ref{rapgen}
with only the first term.
Additional terms would be necessary if we want to include higher order
perturbation terms in Eq.~\ref{pertw} or replace the 
simple MAL
non perturbative part of the 
Wilson loop by more complicate expressions, like that of 
Eq.~\ref{dueventi}. 
Of the terms in $
\lambda$ the $i=j$ terms correspond to self--energies of the quark or of 
the antiquark, the $i \neq j$ terms to the interaction between the quark 
and the antiquark.

   Introducing  Eqs.~\ref{pertw}, 24 in 17,16 one finds 
   explicitly 
    in the momentum space (after factorizing the conservation delta
$(2 \pi)^4  \delta ^4 (p_1 + p_2 - p_1 ^\prime - p_2 ^\prime)$)
\begin{eqnarray}
   & &   \hat{I} (p_1,\,p_2;\,p_1^\prime,\,p_2^\prime)\,=\,
   \hat{I}_{\rm pert} +\hat{I}_{\rm conf} =  16 \pi {4 \over 3} \alpha_s
   \{ D_{\rho \sigma}(Q) q_1^\rho
	q_2^\sigma -\nonumber \\ 
      & & \quad  -{i \over 4} \sigma_1^{\mu \nu} (\delta_\mu^\rho Q_\nu-
     \delta_\nu^\rho Q_\mu) q_2^\sigma D_{\rho \sigma } (Q)
    +{i \over 4} \sigma_2^{\mu \nu} 
	(\delta_\mu^\sigma Q_\nu - \delta_\nu^\sigma Q_\mu) q_1^\rho
  D_{\rho \sigma }(Q)+\nonumber \\
& &
\quad 
 +{1 \over 16} \sigma_1^{\mu_1 \nu_1} \sigma_2^{\mu_2 \nu_2}
	(\delta_{\mu_1}^\rho Q_{\nu_1} - \delta_{\nu_1}^\rho Q_{\mu_1})  
	(\delta_{\mu_2}^\sigma Q_{\nu_2} - \delta_{\nu_2}^\sigma 
Q_{\mu_2})  
	D_{\rho \sigma}(Q) \}  +\ \nonumber \\
& &\quad      
+ \int d^3 {\bf r} \, e^{i {\bf Q} \cdot {\bf r}}\, 
     J({\bf r}, \, q_1, \, q_2) ,
\label{bform} 
\end{eqnarray} 
with
 \begin{eqnarray}
 & &  J({\bf r}, \, q_1, \, q_2)  = {2 \sigma r \over q_{10} + q_{20} }
    \Big [  q_{20}^2 \sqrt{q_{10}^2  -{\bf q}_{\rm T}^2} +  
      q_{10}^2 \sqrt  {q_{20}^2 - {\bf q}_{\rm T}^2} + \label{iconfj} \\
     & &\quad + {q_{10}^2 q_{20}^2 \over \vert {\bf q}_{\rm T} \vert }
      (\arcsin {\vert {\bf q}_{\rm T}\vert \over  q_{10} } + 
      \arcsin {\vert {\bf q}_{\rm T}\vert \over  q_{20}  })
\Big ] +
 {2 \sigma r^k\over r}[  
{\sigma^{k \nu}_1 q_{20} q_{1 \nu} \over  \sqrt{q_{10}^2 
-
 {\bf q}_{ {\rm T}}^2}} - 
 { \sigma^{k \nu}_2 q_{10} q_{2 \nu} 
\over   \sqrt{q_{20}^2 - {\bf q}^2_{ {\rm T}}} }]
+\dots \nonumber
\end{eqnarray}
In such equations $\alpha_s={g^2\over 4 \pi} $ 
denotes  the strong coupling constant and 
$D_{\rho \sigma}(Q)$  the  free  gluon 
propagator; furthermore  we set
 $q_1={p_1 + p_1^\prime \over 2}, q_2={p_2 
      + p_2^\prime \over 2},  Q = p_1 - p_1^\prime   = p_2^\prime - p_2$  
 with $ {\bf q}_1 = -{\bf q}_2 = {\bf q} \, , \ q_{\rm T}^h = 
(\delta^{hk} - \hat r ^h \hat r ^k) q^k $ (CM frame has been assumed).

\subsection{Effective Mass Operator}

    In terms of the relative and the total momentum the homogeneous equation
corresponding to Eq.~\ref{bsh} can be written
\begin{equation}
     \Phi_B (k) = -i \int {d^4 k^\prime
 \over (2 \pi)^4} \hat H_2^{(1)}(\eta_1 P_B
+ k) \hat H_2^{(2)}(\eta_2 P_B - k) \hat I(k , k^\prime;P_B) \Phi_B 
(k^\prime)  ,
\label{onda}
\end{equation}
with $\eta_j={m_j\over (m_1+m_2)}; P_B=(m_B, {\bf 0}) $,
 $m_{\rm B}$ being the mass of the bound state and $\Phi _{\rm B}$
an appropriate wave function. In the 
instantantaneous approximation (consisting in setting $ 
H_j(p)= - {i\over p^2-m_j^2} $ and 
$ k_0=k_0^\prime=\eta_2 {w_1+w_1^\prime \over 2} -
\eta_1 {w_2 +w_2^\prime \over 2}$ in 
$ \hat{I}(k, k^\prime; P) $
with $w_j=\sqrt{m_j^2+{\bf k}^2}$, $w_j^\prime=\sqrt{m_j^2+{\bf k}^\prime}$)
the residual  variables $k_0$ and $k_0^\prime$ can be 
integrated explicitly and one is left with 
  the eigenvalue equation for an  effective mass squared  operator
$ M^2 =M_0^2 + U$ with $ M_0 = w_1 + w_2 $ and
\begin{equation}
  \langle {\bf k} \vert U \vert {\bf k}^\prime \rangle = 
{1\over (2 \pi)^3 } 
 \sqrt{ w_1
  + w_2 \over 2  w_1  w_2}\, \hat I_{\rm inst}(
{\bf k} , {\bf k}^\prime)  \sqrt{ w_1^\prime + w_2^\prime \over 2 
 w_1^\prime w_2^\prime}.
\label{eq:quadrrel}
\end{equation}
From this one can also derive the more conventional operator
 $ M=M_0 +V$, where
\begin{equation}
\langle {\bf k} \vert V \vert {\bf k}^\prime \rangle =
 ={1 \over ( 2 \pi)^3 }
 {1 \over 4 \sqrt{ w_1 w_2 w_1^\prime w_2^\prime } } \hat{I}_{\rm inst} 
({\bf k}, {\bf k}^\prime)+\dots 
\label{linrel}
\end{equation}
having kept only first order terms in $U$ and neglected kinematical factors 
 equal  to 1 on the energy shell $(w_1+w_2=w_1 ^\prime +w_2 ^\prime)$.

     From Eqs.~\ref{bform}-- \ref{iconfj} one obtains explicitly
\begin{eqnarray}
& & \langle {\bf k} \vert U \vert {\bf k}^\prime \rangle =
\sqrt{(w_1+w_2) (w_1^\prime +w_2^\prime)
\over w_1 w_2 w_1^\prime w_2^\prime}
\bigg \{
- {4\over 3} {\alpha_s \over \pi^2}  \Big [ {1\over {\bf Q}^2}
\big [ q_{10} q_{20} + {\bf q}^2 + { ( {\bf Q}\cdot {\bf q})^2 \over
{\bf Q}^2 } ] \nonumber \\
& & + {i\over 2 {\bf Q}^2} {\bf k}\times {\bf k}^\prime \cdot ({\bf \sigma}_1 +
{\bf \sigma}_2 ) + {1\over 2 {\bf Q}^2 } [ q_{20}
(\alpha_1 \cdot {\bf Q}) - q_{10} (\alpha_2\cdot {\bf Q}) ]+
\nonumber \\
& & + {1\over 6} {\bf \sigma}_1 \cdot {\bf \sigma}_2 + {1\over 4} 
( {1\over 3} {\bf \sigma}_1 \cdot {\bf \sigma}_2 - { ( {\bf Q}\cdot \sigma_1)
( {\bf Q}\cdot {\bf \sigma}_2) \over {\bf Q}^2 })
+ {1\over 4 {\bf Q}^2 } ( \alpha_1 \cdot {\bf Q}) ( \alpha_2 \cdot {\bf Q})
\Big ]
\label{uconf}\\
& &
+{1\over ( 2 \pi)^3} \int d^3{\bf r} e^{i {\bf Q}\cdot {\bf r}}
J^{\rm inst}({\bf r}, {\bf q}, q_{10}, q_{20})\bigg \} \nonumber
\end{eqnarray}
with
\begin{eqnarray}
& & 
J^{\rm inst}({\bf r}, {\bf q}, q_{10}, q_{20})= {\sigma r \over q_{10}+q_{20}}
[ q_{20}^2 \sqrt{q_{10}^2-{\bf q}^2_t} + q_{10}^2 \sqrt{q_{20}
-{\bf q}_{\rm T}^2}
] +\nonumber\\
& & \quad + {q_{10}^2 q_{20}^2 \over \vert {\bf q}_{\rm T} \vert}
(\arcsin{\vert{\bf q}_{\rm T}\vert \over q_{10} } 
+ \arcsin{\vert {\bf q}_{\rm T}\vert
\over q_{20}} )]\label{uconf1}\\
& & \quad - {\sigma\over r} [ {q_{20} \over \sqrt{q_{10}^2-
{\bf q}^2_{\rm T}}}
( {\bf r} \times {\bf q}\cdot \sigma_1 + i q_{10} ({\bf r}\cdot \alpha_1))
+ {q_{10} \over \sqrt{q_{20}^2 - {\bf q}^2_{\rm T}}} 
( {\bf r}\times {\bf q} \cdot
\sigma_2 - i q_{20} ( {\bf r}\cdot{\bf \alpha}_2))]\nonumber
\end{eqnarray}
Here $ \alpha_j^k $ denote the usual Dirac matrices $\gamma_j^0 \gamma_j^k$,
$\sigma_j^k$ the $4 \times 4$ Pauli matrices $ \left( \matrix {\sigma_j^k &
0 \cr 0 & \sigma_j^k}\right ) $ and obviously
$
{\bf q} ={ {\bf k}+ {\bf k}^\prime \over 2}, \quad \quad {\bf Q}= {\bf k}
- {\bf k}^\prime , \quad q_{j0}= {w_j+w_j^\prime \over 2}
$.
Notice that, due to the terms in $ \alpha_j^k $, such $U$ is hermitian
only with reference to the undefined metric operator $\gamma_1^0 \gamma_2^0$. 

   Due to
 Eq.~\ref{linrel} the potential $V$ can be obtained from $U$ as given by
Eqs.~\ref{uconf}-\ref{uconf1} simply by the kinematical replacement
$
\sqrt{ (w_1+w_2) (w_1^\prime +w_2^\prime)\over w_1w_2w_1^\prime w_2^\prime}
\to {1\over 2 \sqrt{w_1 w_2 w_1^\prime w_2^\prime}}
$.
Notice that, if in the resulting expression we perform an $1/m$ expansion and 
make an appropriate Foldy-Wouthuysen transformation to eliminate the terms in
$\alpha_j^k$, we reobtain \cite{bsnoi} the MAL potential as given 
in Tab. 1.

  On the contrary, if we keep only the long range terms in $V$, neglect the 
spin dependent part and set for simplicity $m_1 = m_2 = m $, we can write in
an operatorial form
\begin{equation}
M= 2 \sqrt{m^2 + {\bf q}^2} + {\sigma r\over 2} 
\big [ {\sqrt{m^2 + {\bf q}^2}\over \vert {\bf q}_{\rm T}\vert } 
\arcsin { \vert {\bf q}_{\rm T} \vert \over \sqrt{m^2 + {\bf q}^2}}
+ \sqrt{m^2 + {\bf q}^2_{\rm L} \over m^2 + {\bf q}^2_{\rm T}}]
\label{eqrelols}
\end{equation}
(${\bf q}_{\rm L}={\bf q} -{\bf q}_{\rm T}$)
in which the ordering implied by the definition
referring to Eq.~\ref{uconf1} has to be understood. Such 
expression is identical to the hamiltonian of the Relativistic Flux Tube Model
\cite{olsson}, \cite{fluxnoi}
 up to the order $\sigma a^2 $. Two different limits
 of~\ref{eqrelols}
are of interest. The first one is the limit of small angular momentum. i.e. 
small transversal momentum ($q_{\rm T}^2 = {L^2 \over r^2}$).
 In this case we have (strictly for $s$ waves)
\begin{equation}
M= \sqrt{m^2 + {\bf q}^2} + \sigma r
\label{relrad} 
\end{equation}
This result justifies the use that of Eq.~\ref{relrad} has been done in 
the study of light meson spectrum. The second limit
 is for large angular momentum
or transversal momentum (negligible $ q_{\rm L}^2$) and is
\begin{equation}
M= 2 \sqrt{m^2 +{\bf q}^2} + {\sigma r\over 2} ( {\sqrt{m^2+{\bf q}^2}
\over \vert {\bf q}\vert } \arcsin {\vert {\bf q}\vert \over 
\sqrt{m^2+{\bf q}^2 }} + {m\over \sqrt{m^2 +{\bf q}^2}})
\end{equation}
or under the additional assumption of negligible $m$ 
\begin{equation}
M= 2 \vert {\bf q} \vert + {\pi \over 4 }\sigma r
\label{imp}
\end{equation}
Eq.~\ref{imp} is very important for an understanding of the Regge trajectory
properties. In fact it is well known that such hamiltonian produces asympotically
straight  trajectories with
\begin{equation}        
 m_l^2 \rightarrow 8{\pi \over 2} \sigma l = 2 \pi \sigma l 
\label{numquant}
\end{equation}
$m_l$ being obviously the mass of  a given radial quantum 
number  bound state  as a function of 
the angular momentum $l$. Notice that Eq.~\ref{numquant} corresponds to the 
slope
$
\alpha^\prime = {dl \over d m^2 } \to {1\over 2 \pi \sigma }$
which is identical to that of the Nambu-Goto string model. Notice that 
for $\sigma = 0.18$ we find $\alpha^\prime=0.88$ in very good agreement with
the the experimental slope of the $\rho$ trajectory. Had we used the naive
 Eq.~\ref{relrad}, 
even for large $l$, we would have find $ \alpha^\prime = 1/8 \sigma = 0.69$. 
As mentioned
such results are  in perfect agreement  with 
 those of Refs.\cite{Dubin},  \cite{olsson}.

   Finally, if we keep all terms in Eq.~\ref{uconf}, but assume
$m_2 \gg m_1$, we can discuss the light--heavy 
quark symmetry much on the lines 
followed by Kaidalov \cite{kaidalov}. E. g. let us consider the   
quadratic hyperfine separation between triplet and singlet $ \delta m^2 =
m_1^2-m_2^2 $ in the (uQ) states. Empirically this
quantity is nearly independent of Q ($\sim 0.55
\quad {\rm MeV}^2 $). In fact the hyperfine splitting term in $U$
(cf.~\ref{uconf})
depends on the quark masses only through the kinematical
factor occurring in it. Now for $m_2 \gg m_1$ such factor reduces to
$ \sqrt{1 \over w_1 w_1^\prime }$ and any dependence on $m_2$ disappears.

 \section*{Acknowledgments}
 We gratefully acknowledge discussions with Baker, Dosch, Dubin, 
 Kaidalov, Simonov.

 \end{document}